\documentclass[aps,pre,final,letterpaper,10pt,twocolumn,floats,showpacs,amsmath,amsfonts,amssymb]{revtex4-2}
\usepackage{graphicx} 
\usepackage{epsf}
\usepackage{epstopdf}
\usepackage{tikz}
\usepackage{float}
\usepackage[colorlinks=true,citecolor=blue]{hyperref}
\usepackage[normalem]{ulem}

\begin{document}

\title{Formation of Fano resonance in double quantum dot system}

\author{Grzegorz Micha{\l}ek} \email{grzechal@ifmpan.poznan.pl}
\author{Bogdan R. Bu{\l}ka}

\affiliation{Institute of Molecular Physics, Polish Academy of Sciences, ul. Mariana Smoluchowskiego 17, 60-179 Pozna\'{n}, Poland}

\begin{abstract}
Transient electron dynamics and the Fano resonance formation over time in transport through a two quantum dot system in a T-shape geometry (2QD-T) are analysed for free electrons.
Time evolution of the transport characteristics are different for a weak and a strong inter-dot coupling with respect to the dot-electrode coupling.
When the dot-dot coupling is weak, we find that transport dynamics is governed by two channel tunneling processes with two different relaxation times and buildup of the Fano resonance is developed much slower, with the longer relaxation time.
In the opposite limit, the transmission evolves with a single relaxation time, from the single peak resonance structure into the structure with two asymmetric peaks.
One observes large dot-electrode charge oscillations, which can temporarily change direction of transient currents.
We find also that the time evolution of entanglement is different in both coupling regimes; in particular rapid Rabi oscillations caused by coherent electron transfer between the 2QD states can be seen for the strong dot-dot coupling.
\end{abstract}

\maketitle

\section{Introduction}

Nowadays, one can experimentally observe dynamics of charge flow in nanostructures under the non-stationary conditions, after applying short-time voltage pulses or other sudden changes in system parameters.
The ultra-fast time-re-solved spectroscopy techniques enable real-time observation of, among others, how the system returns to equilibrium from excited states.
Just recently the Fano resonance buildups were monitored in optical experiments using the spectrally resolved electron interferometry and the atto-second transient absorption spectroscopy~\cite{gruson2016, kaldun2016}.
The Fano resonance was observed in electron transport measurements in nanostructures, namely in the system of a strongly coupled quantum dot with electrodes \cite{gores2000, sasaki2009}.
Two quantum dots in a T-shape geometry (2QD-T) seems to be a suitable system for theoretical studies of the Fano resonance, because conductance shows a characteristic asymmetric dip due to destructive interference of a travelling wave through the central QD with an localized state at the side attached QD ~\cite{bulka2005, zitko2010} (see also~\cite{bulka2001, kang2001, stefanski2004} and references therein).

However the analysis of transient  dynamics in the 2QD-T system has been limited to coherent ringing (oscillations) or beats in transient currents after a switching on or off a bias voltage~\cite{pan2009}.
In Ref. \cite{kwapinski2014} a charging effect and transient currents were studied in a T-shaped wire of $N$ QDs after a sudden change of the bias voltage.
Therefore, in the paper we focus on an unexplored topic:
Fano resonance buildup over time in the 2QD-T system when the second QD is suddenly side attached.
In particular, we will study transient currents and the time evolution of transmission to identify relaxation processes in different time regimes.

\section{Description of DQD system and method of calculations}

We consider two single-level QD in a T-shape geometry coupled with two normal-metal left (L) and right (R) electrodes.
We assume that in a distant past only the central QD$_1$ is coupled with the electrodes under bias and it is in a stationary state, while the second QD$_2$ is suddenly attached to the QD$_1$ at the time $t = 0$.
The Hamiltonian of the system (for the spinless noninteracting electrons) can be written in the form $H = H_{QD} + \sum_{\alpha = L,R} H_\alpha + H_T$, where
\begin{align}
&H_{QD} = \sum_{i = 1, 2} \epsilon_i d_i^\dag d_i + \Theta(t) t_{12} ( d_1^\dag d_2 + d_2^\dag d_1 ) \; , \nonumber \\
&H_\alpha = \sum_k ( \epsilon_{\alpha k} - \mu_\alpha ) c_{\alpha k}^\dag c_{\alpha k} \; , \\
&H_T = \sum_{\alpha k} t_{\alpha} ( d_1^\dag c_{\alpha k} + c_{\alpha k}^\dag d_1 ) \; . \nonumber
\end{align}
Here, $d_i^\dag (d_i)$ stands for creation (annihilation) operator for an electron on the QD$_i$ and at the dot level $\epsilon_i$.
An abrupt coupling of the QD$_2$ at $t = 0$ is described by time-dependent hopping between the dots $\Theta(t) t_{12}$, where $\Theta(t)$ is a Heaviside step function and $t_{12}$ denotes the inter-dot hopping parameter.
$c_{\alpha k}^\dag (c_{\alpha k})$ describes electrons with the energy $\epsilon_{\alpha k}$ in the $\alpha$ electrode which has a chemical potential $\mu_\alpha$.
The coupling of the central QD$_1$ with the electrodes is given by the hopping parameter $t_{\alpha}$.

The time-dependent electrical currents flowing into the QD$_1$ from the $\alpha$ electrode can be calculated from the evolution of the total number operator using the equation of motion technique (EOM) for the non-equilibrium Green function~\cite{haug2008}
\begin{equation}
J_\alpha(t) = - 2 \frac{i e t_\alpha}{\hbar} \Im \sum_k \langle d_1^\dag(t) c_{\alpha k} (t) \rangle \; ,
\end{equation}
where the electrodes are treated in the wide-band approximation.
For the creation (annihilation) operators, in the QDs and the electrodes, we write appropriate Heisenberg equation of motion. Next we solve this system of equations by performing a Laplace transform~\cite{schaller2014}.
The time dependencies of the operators are found by inverting the Laplace transform in the wide band limit~\cite{schaller2014}.
In general, the current $J_L(t)$ can be written in the form:
\begin{equation}
J_L(t) = \frac{e}{h} \int_{-\infty}^{\infty} [ T(E,t) \Delta f(E) + A(E,t) f_W(E) ] dE \; ,
\end{equation}
where $\Delta f(E) \equiv f_L(E) - f_R(E)$, $f_W(E) \equiv [ \Gamma_L f_L(E) + \Gamma_R f_R(E) ] / ( \Gamma_L + \Gamma_R )$ denotes the Wigner distribution function for electrons at the 2QD system and $f_\alpha(E) \equiv f ( E - \mu_\alpha )$ is the Fermi-Dirac distribution in the electrode $\alpha$ with a chemical potential $\mu_\alpha$.
The tunneling rates are $\Gamma_\alpha \equiv 2 \pi \rho_\alpha t_\alpha^2$, where $\rho_\alpha$ is a constant density of state in the electrode $\alpha$.
$T(E,t)$ denotes a transmission evolution, while $A(E,t)$ is related with a local charge and entanglement evolution \cite{amico2008, laflorencie2016} to the stationary state.
Both coefficients include contributions to quantum interference from various tunneling processes.

\section{Results}

To simplify considerations we assume that both QD are identical with $\epsilon_1 = \epsilon_2 = \epsilon$ and there is a symmetrical coupling to the electrodes, $\Gamma_L = \Gamma_R = \Gamma$.
Since time evolution of transport characteristics and electron dynamics are different for a weak and a strong inter-dot coupling $t_{12}$ with respect to $\Gamma$, we present them separately.

\subsection{Weak dot-dot coupling}

\begin{figure}
\centering
\includegraphics[scale=.48]{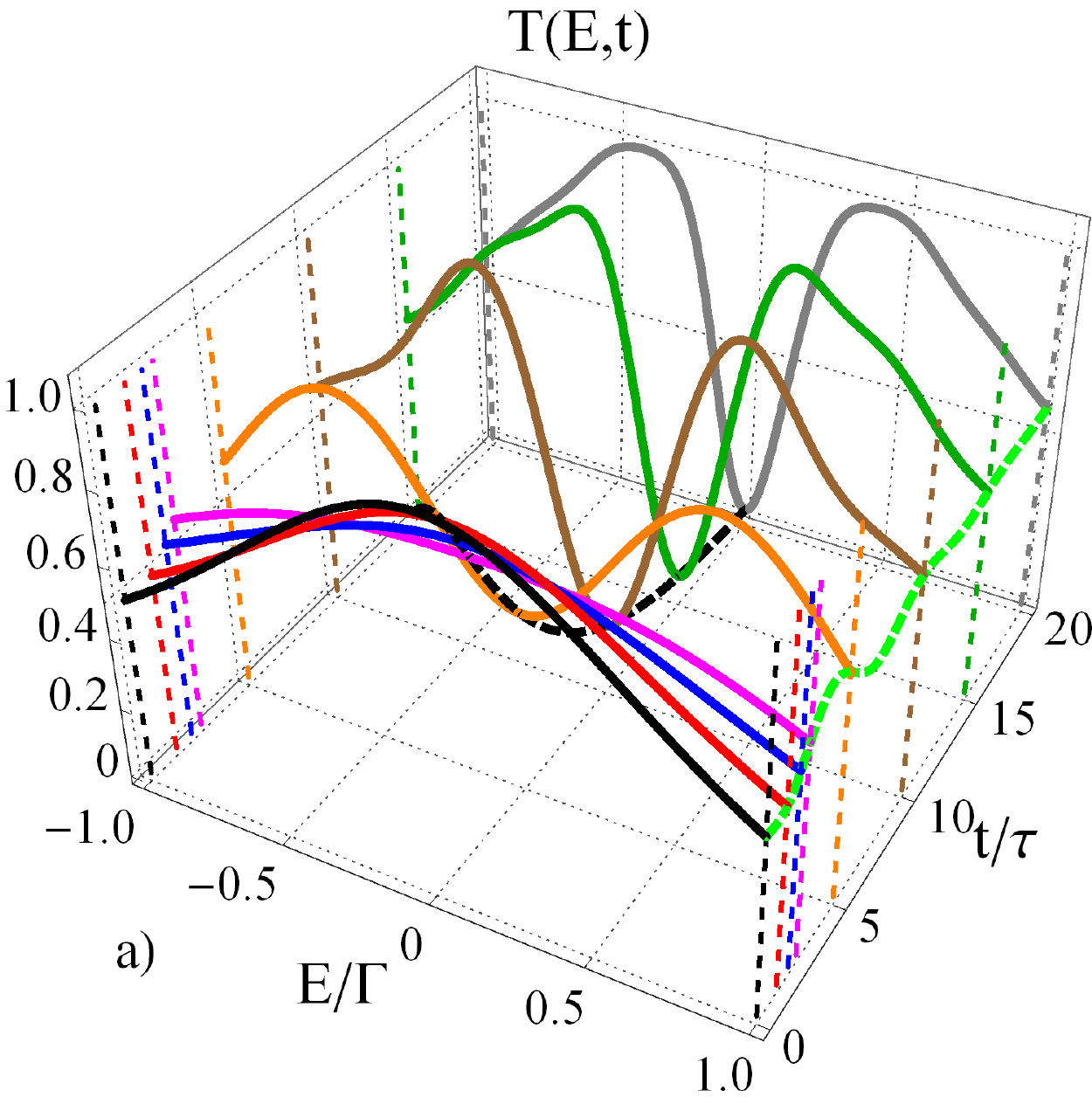} \\
\includegraphics[scale=.48]{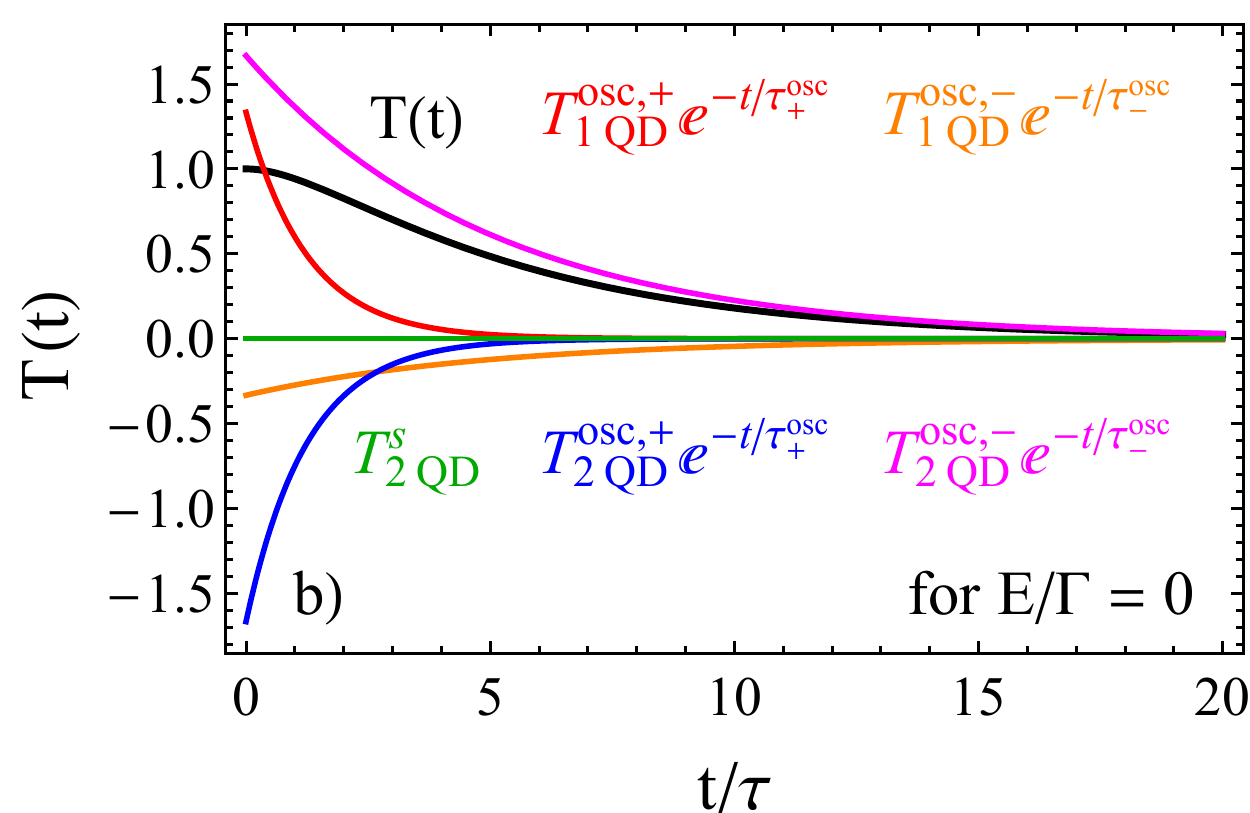} \\
\vspace{-0.8cm}
\includegraphics[scale=.48]{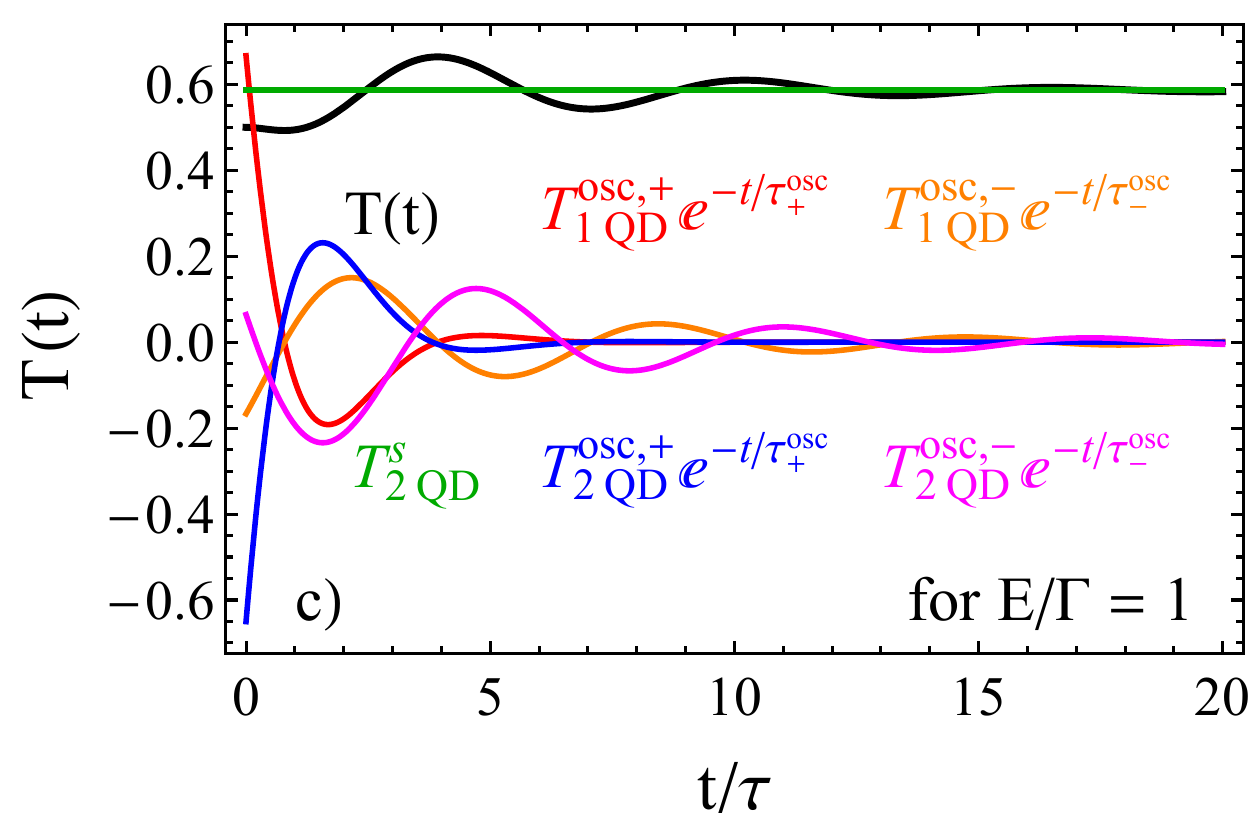}
\caption{a) $T(E,t)$ as a function of energy $E$ for selected times $t = \{ 0, 2 \tau_+, 2 \tau, 2 \tau_+^{osc}, 2 \tau_-, 2 \tau_-^{osc}, 3 \tau_-^{osc}, 4 \tau_-^{osc} \}$ (black, red, blue, magenta, orange, brown, green, gray curves) for small $t_{12} / \Gamma = 0.4$.
b) and c) The time evolution of $T(E,t)$ and its components for $E = 0$ and $E = 1$, respectively.
The other parameters are $\epsilon = 0$, $\Gamma = 1$, $\tau = 1 / \Gamma = 1$.} \label{fig1}
\end{figure}

\begin{figure}
\centering
\includegraphics[scale=.48]{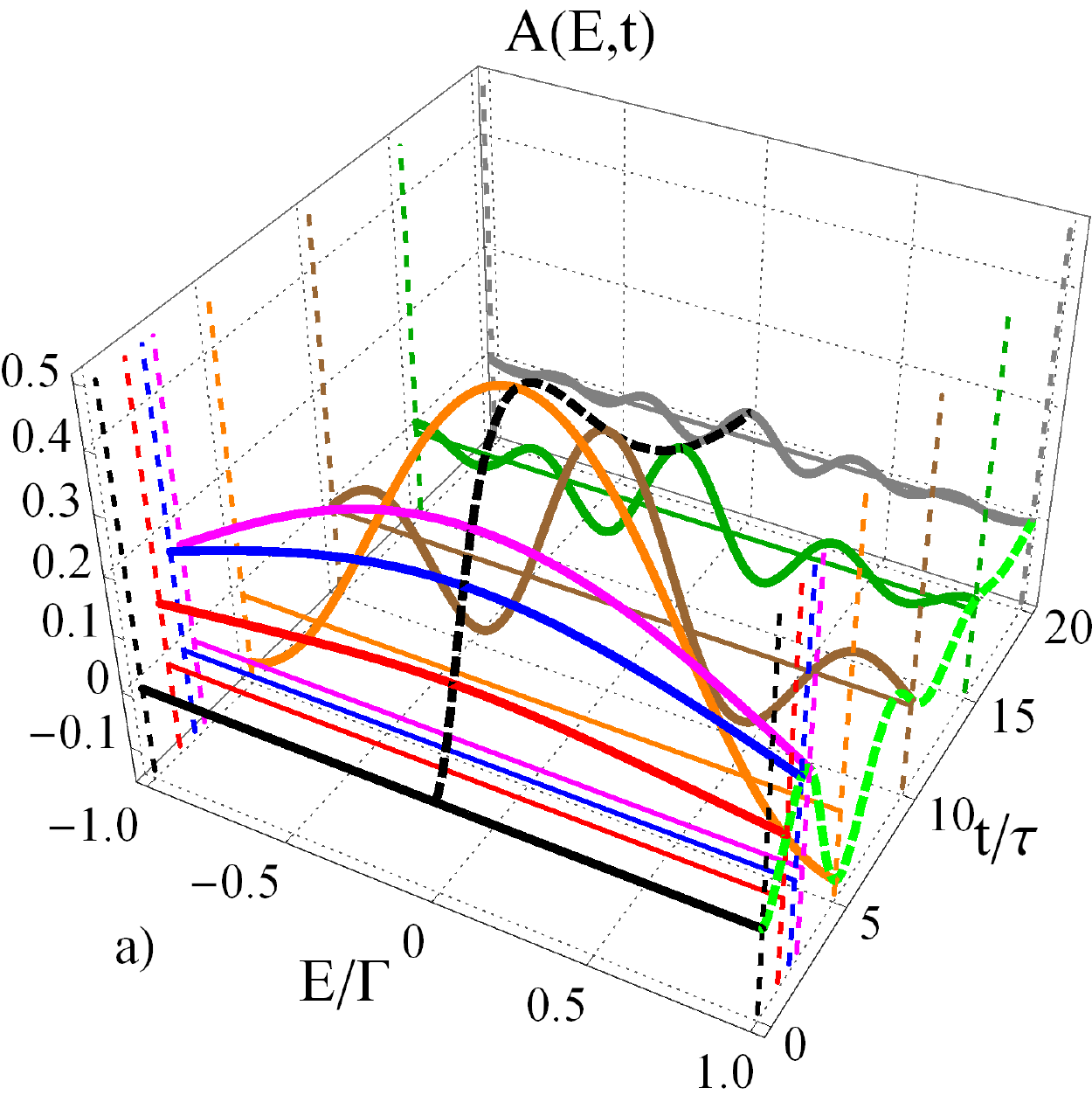} \\
\includegraphics[scale=.48]{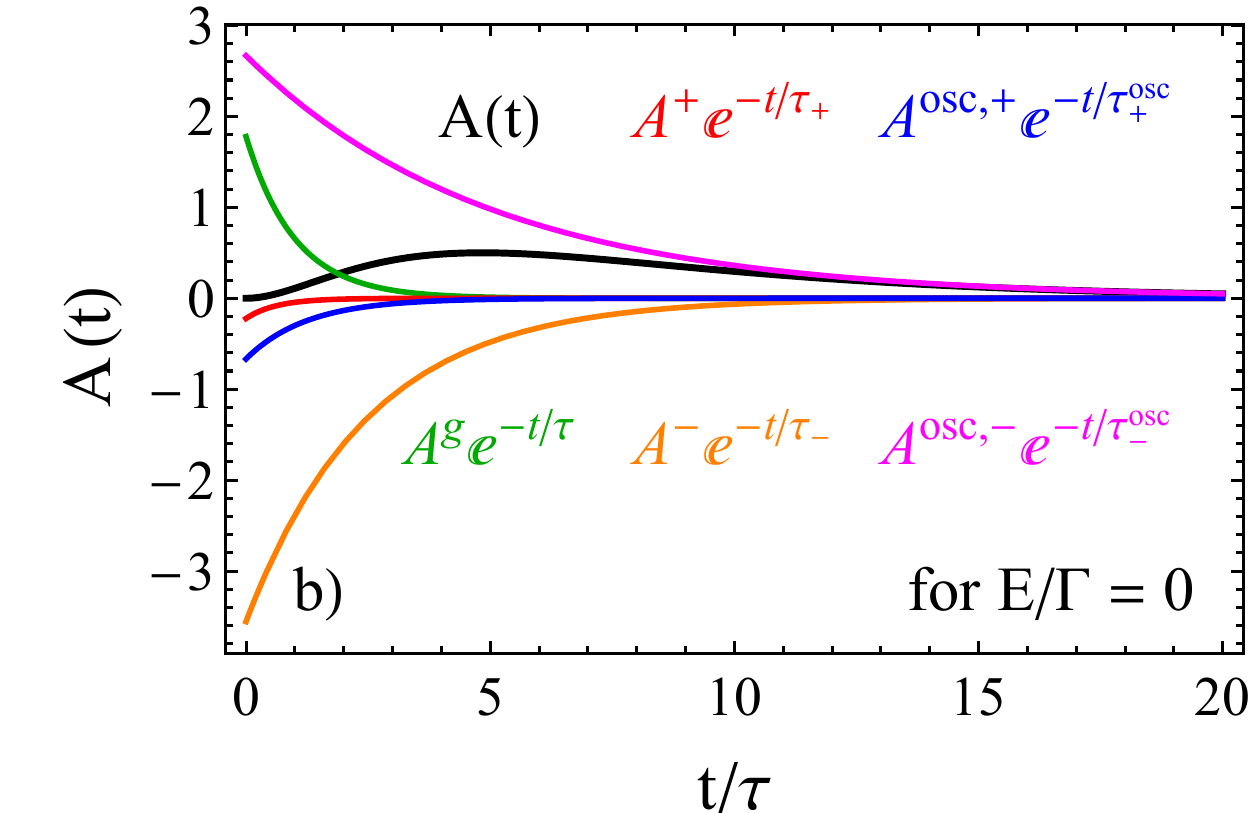} \\
\vspace{-0.8cm}
\includegraphics[scale=.48]{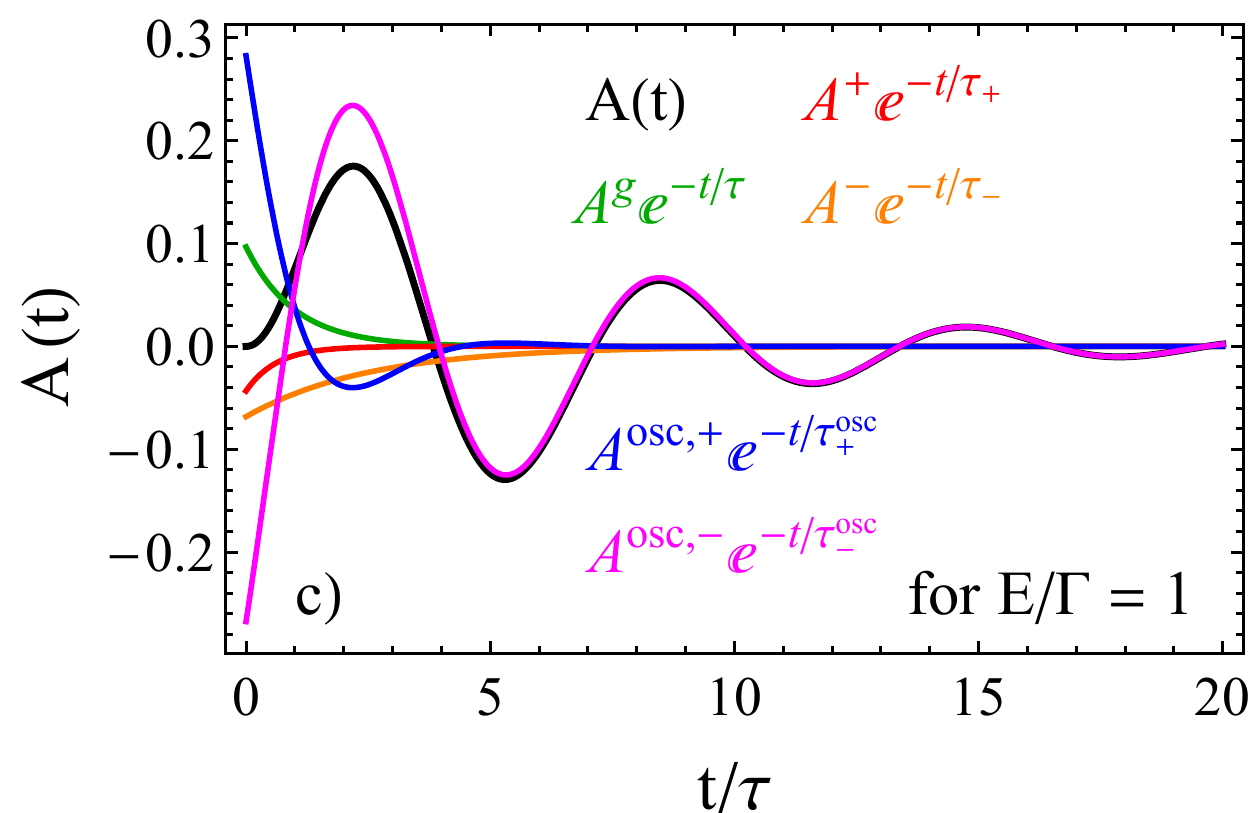}
\caption{a) $A(E,t)$ as a function of energy $E$ for selected times $t$ for small $t_{12} / \Gamma = 0.4$.
b) and c) The time evolution of $A(E,t)$ and its components for $E = 0$ and $E = 1$, respectively.
The other parameters are the same as in Fig.~1.} \label{fig2}
\end{figure}

We start from the weak dot-dot coupling case, i.e. when $t_{12} \ll \Gamma$.
In this case one level is strongly hybridized with the electrodes, while the other one (on the QD$_2$) is localized and only weakly coupled to a conducting channel.
Interference between these two conducting channels leads to a Fano anti-resonance with a characteristic dip in the transmission.
The time evolution of the transmission can be expressed as
\begin{align}\label{eq:TEt}
T(E,t) &= \sum_{\nu = \pm} [ T_{1QD}^{osc,\nu}(E,t) - T_{2QD}^{osc,\nu}(E,t) ] e^{-t / \tau_\nu^{osc}} \nonumber \\
&+ T_{2QD}^s(E) \; .
\end{align}
The components
\begin{align}
&T_{1QD}^{osc,\nu}(E,t) = \frac{\nu \Gamma \Gamma_\nu}{2 \delta} \frac{\Gamma \cos ( \mathcal{E} t ) - \mathcal{E} \sin ( \mathcal{E} t )}{\mathcal{E}^2 + \Gamma^2} \; , \\
&T_{2QD}^{osc,\nu}(E,t) = \frac{\nu \Gamma \Gamma_\nu}{4 \delta} \frac{\Gamma_\nu \cos ( \mathcal{E} t ) - 2 \mathcal{E} \sin ( \mathcal{E} t )} {\mathcal{E}^2 + \Gamma_\nu^2 / 4} \;
\end{align}
and they describe a time evolution of the transmission related with the 1QD system and the creation of the Fano resonance in the 2QD system with the same two characteristic times $\tau_\pm^{osc} \equiv 2 / \Gamma_\pm$, where $\Gamma_\pm = \Gamma \pm \delta$ and $\delta = ( \Gamma^2 - 4 t_{12}^2 )^{1/2}$.
In this case the relaxation processes depend not only on the coupling $\Gamma$ with the external electrodes but they are also strongly modified by the hopping $t_{12}$ between QDs, which significantly changes transport properties of the system.
Here, we denoted by $\mathcal{E} = E - \epsilon$, and took $e \equiv 1$ and $h \equiv 1$.
Above we performed the spectral decomposition to show a role of both relaxation processes with their characteristic times $\tau_\pm^{osc} = 2 / \Gamma_\pm$.
The last term corresponds to the transmission in the stationary state (at $t = \infty$) and after the spectral decomposition it can be expressed as $T_{2QD}^s(E) = \sum_{\nu = \pm} T_{2QD}^{s,\nu}(E)$, where
\begin{equation}\label{eq:t2qds}
T_{2QD}^{s,\nu}(E) = \frac{ \Gamma}{4 \delta} \frac{\nu \Gamma_\nu^2}{\mathcal{E}^2 + \Gamma_\nu^2 / 4} \; .
\end{equation}
This result can be interpreted as resonant transport through two conducting channels with two different broadenings $\Gamma_\pm / 2$. Notice that $T_{2QD}^{s,-}(E)$ is negative and describes the destructive interference.

Fig.~1a) presents a 3D plot of $T(E,t)$, the function which is always positive, although its components can be positive as well as negative.
Fig.~1b) shows the evolution of the Fano dip (at $E = 0$).
In this case $T_{2QD}^s(0) = 0$, the components $T_{1QD}^{osc,\pm}(0) = \pm \Gamma_\pm / ( 2 \delta )$, $T_{2QD}^{osc,\pm}(0) = \pm \Gamma / \delta$ and thus, $T(0,t) = ( - \Gamma_- e^{- t / \tau_+^{osc}} + \Gamma_+ e^{- t / \tau_-^{osc}} ) / ( 2 \delta )$ vanishes monotonically.
One can say that the Fano dip evolves with both relaxation times $\tau_\pm$, but destructive interference dominates in a long time limit, for $t > \tau_-^{osc}$.
For $\mathcal{E} \neq 0$ one finds oscillations in $T(E, t)$ with the period $\omega = 2 \pi |\mathcal{E}|$.
These oscillations are related to coherent charge transfers between the QD system and the electrodes.
In general, these oscillations are dominated by $T_{1QD}^-(E,t)$ and $T_{2QD}^-(E,t)$ component, while $T_{1QD}^+(E,t)$ and $T_{2QD}^+(E,t)$ component become relevant only in the short time regime.

The spectral function $A(E,t)$ describes processes associated with establishing a local chemical potential (related with the Wigner function) and with changes of entanglement in the system, also in presence of the bias voltage.
It can be written in the form
\begin{align}
\label{eq:A}
&A(E,t) = A^g(E) e^{- t / \tau} \nonumber \\
&+ \sum_{\nu = \pm} [ A^\nu(E) e^{- t / \tau_\nu} + A^{osc,\nu}(E,t) e^{- t / \tau_\nu^{osc}} ] \; ,
\end{align}
where components $A^g(E)$, $A^\nu(E)$ and $A^{osc,\nu}(E,t)$ are complex and are given in \textit{Appendix}.
$A(E,t)$ vanishes in the stationary states, i.e. for $t = 0$ or $t = \infty$ (see Fig.~2), because we consider symmetric coupling with the electrodes when local charge accumulation is absent.
One can see that in $A(E,t)$ new terms appear which are related with additional relaxation processes (with a relaxation time $\tau \equiv 1 / \Gamma$ and $\tau_\pm \equiv 1 / \Gamma_\pm$).
They are relevant only for short times ($t < \tau_-^{osc}$), while for larger ones $A^{osc,-}(E,t)$ component plays a dominant role.
In contrast to $T(E,t)$, energy-dependent oscillations are clearly visible in $A(E, t)$.
Their period increase with time.
Let us take a look at $E = 0$.
$A(0,t)$ is positive but a non-monotonic function of time due to an interplay of the components $A^g(0) = \Gamma_+ \Gamma_- / \delta^2$, $A^\pm(0) = - \Gamma_\mp^2 / (2 \delta^2 )$ and $A^{osc,\pm}(0) = \mp \Gamma_\mp / \delta$ which vanish with the relaxation times $\tau$, $\tau_\pm$ and $\tau_\pm^{osc}$, respectively.

\subsection{Strong dot-dot coupling}

\begin{figure}
\centering
\includegraphics[scale=.48]{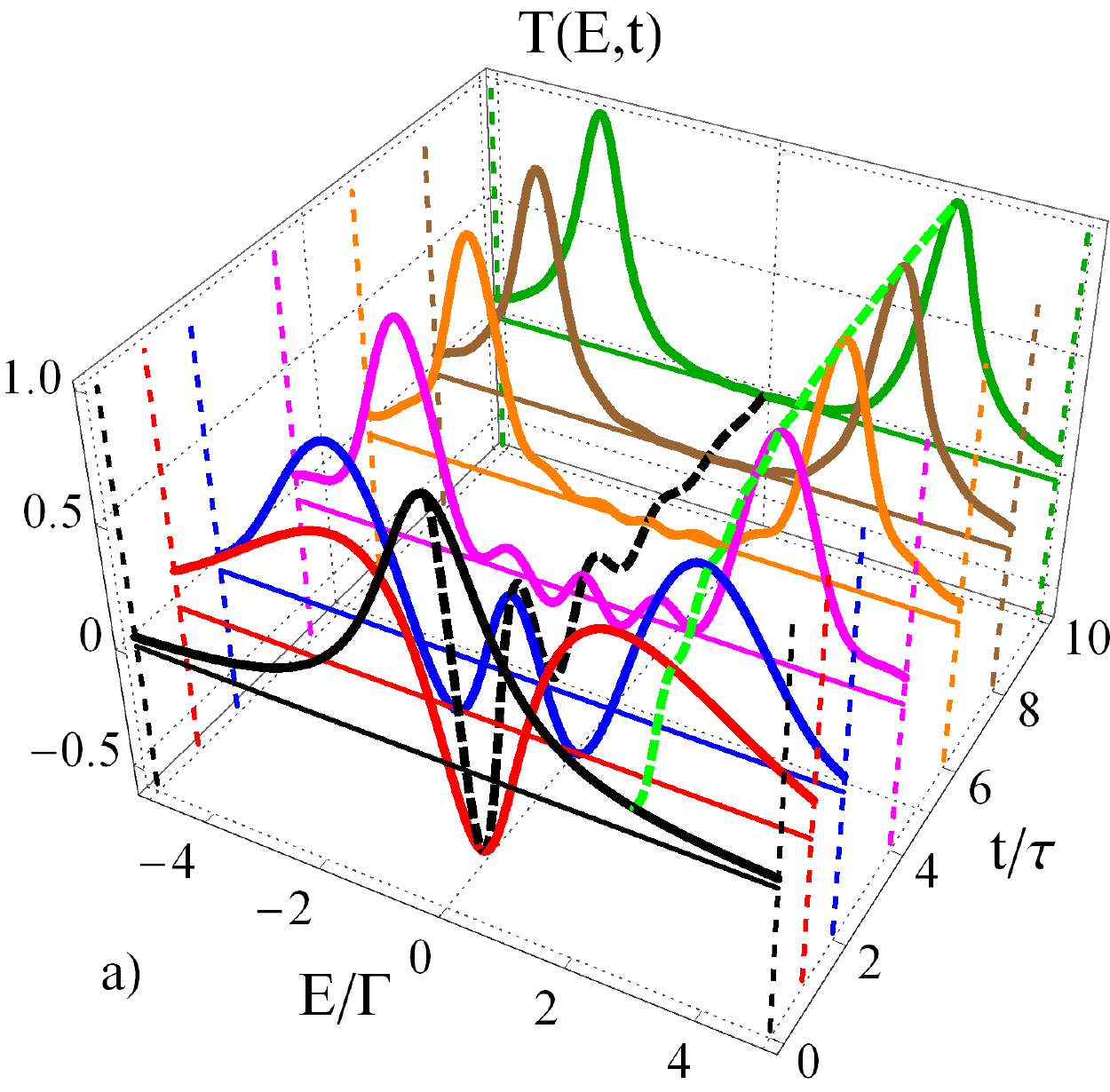} \\
\includegraphics[scale=.48]{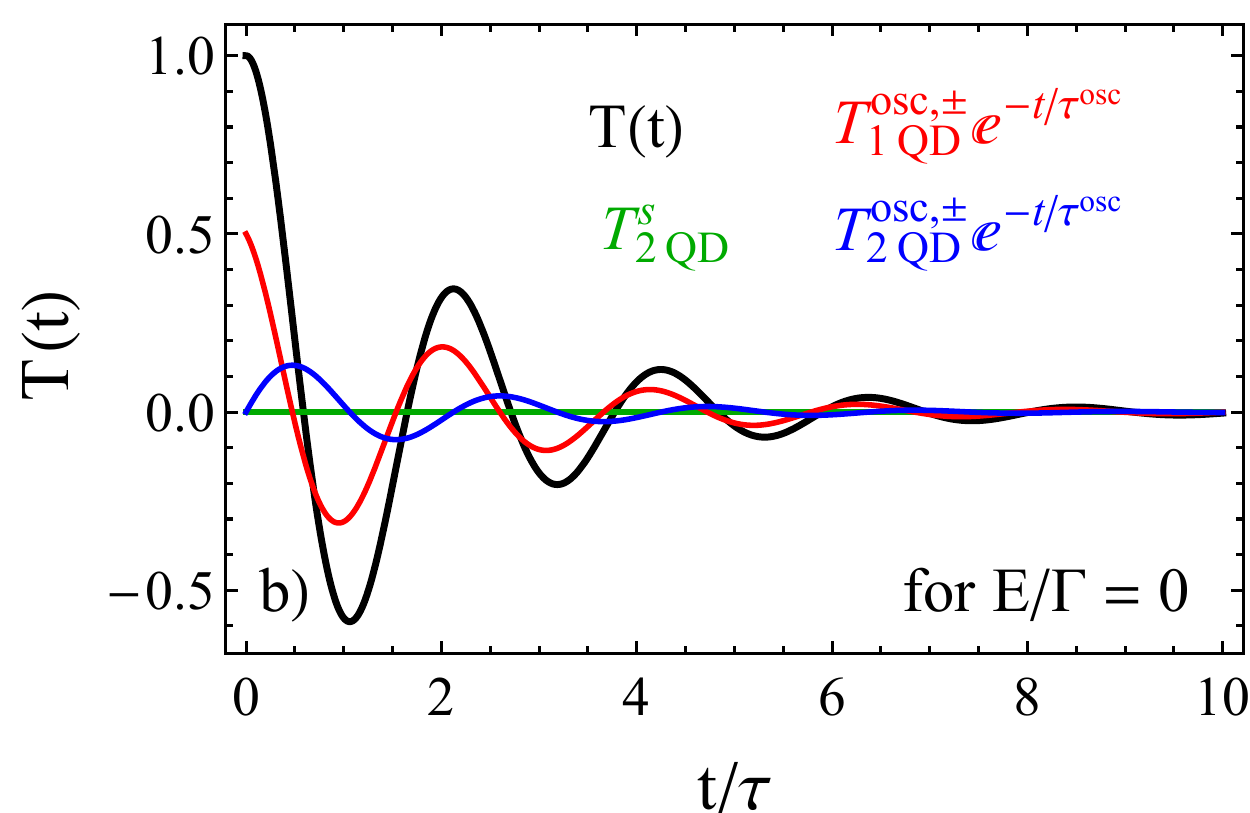} \\
\vspace{-0.8cm}
\includegraphics[scale=.48]{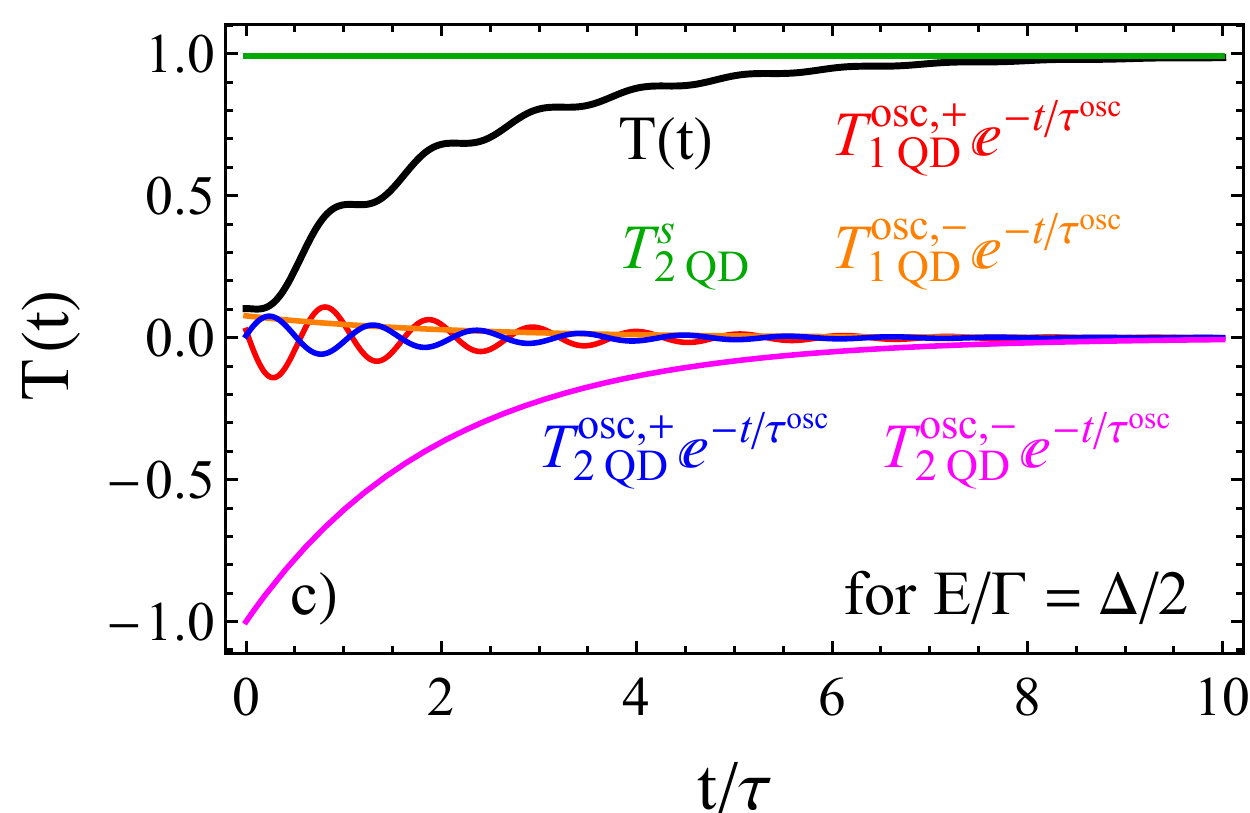}
\caption{a) $T(E,t)$ as a function of energy $E$ for selected $t / \tau^{osc} = \{ 0, 1, 2, 4, 6, 8, 10 \}$ (black, red, blue, magenta, orange, brown, green curves) for large $t_{12} / \Gamma = 3$.
b) and c) The vanishing of the 1QD peak and formation of the anti-bonding state for $E = 0$ and $E = E_-$, respectively.
The other parameters are $\epsilon = 0$, $\Gamma = 1$ and $\tau = 1 / \Gamma$.} \label{fig3}
\end{figure}

\begin{figure}
\centering
\includegraphics[scale=.48]{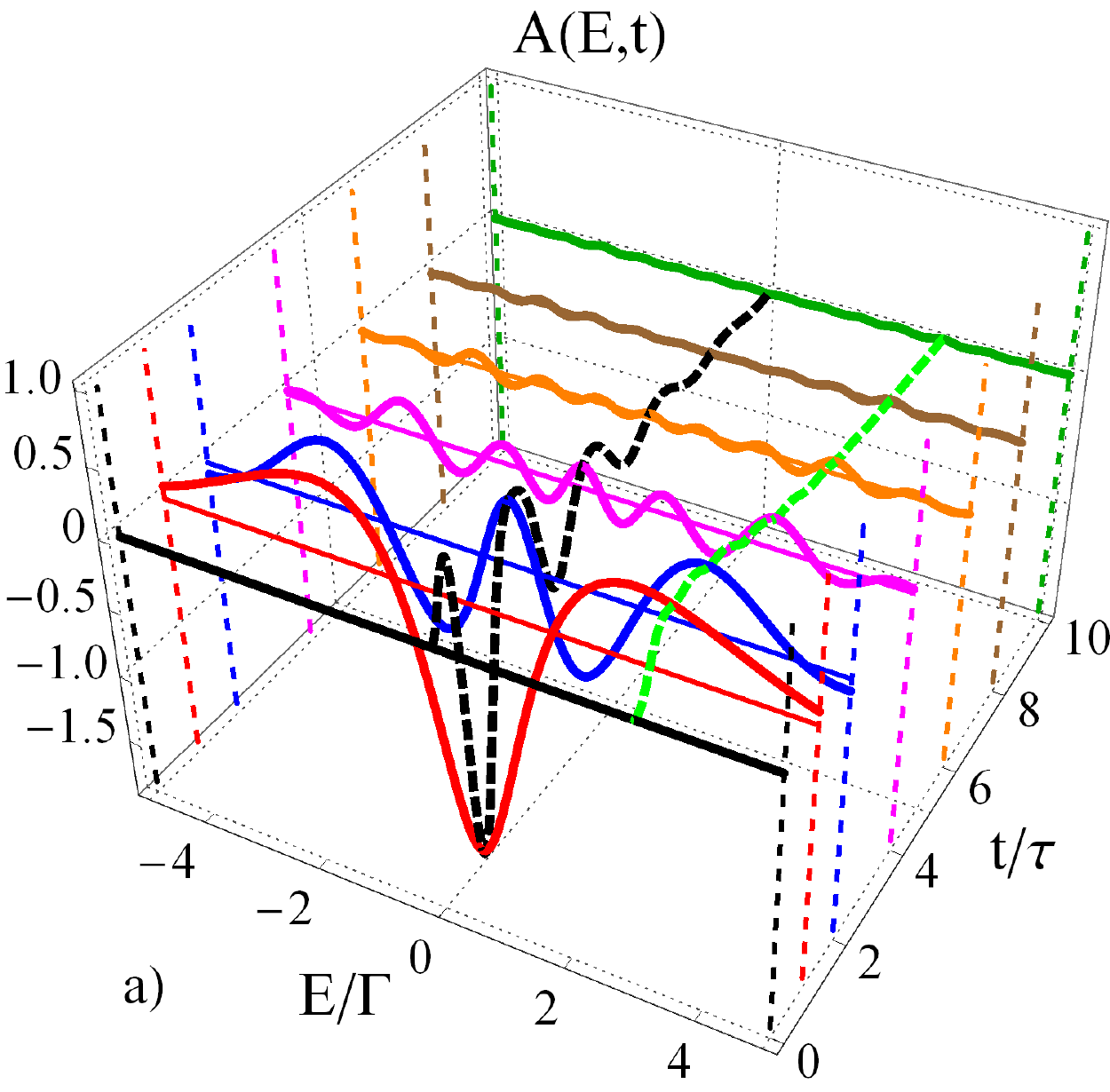} \\
\includegraphics[scale=.48]{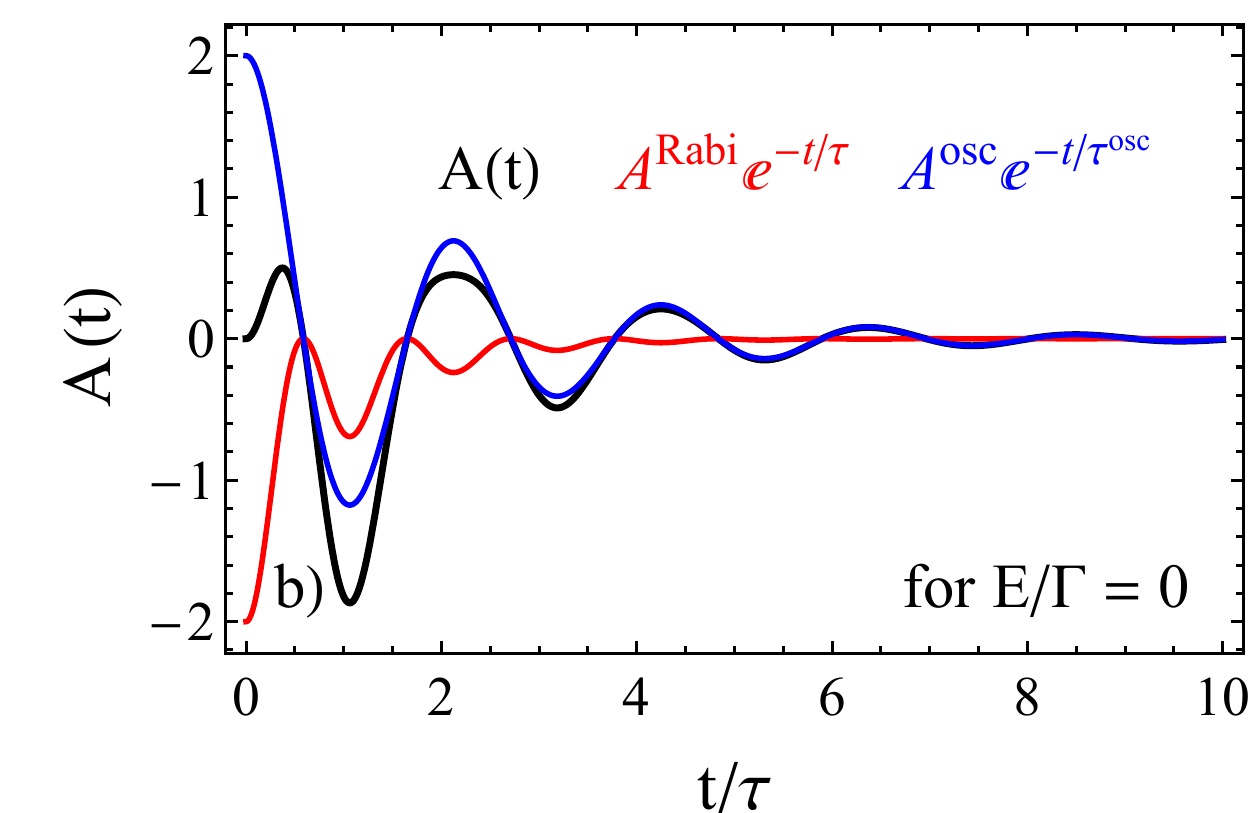} \\
\vspace{-0.8cm}
\includegraphics[scale=.48]{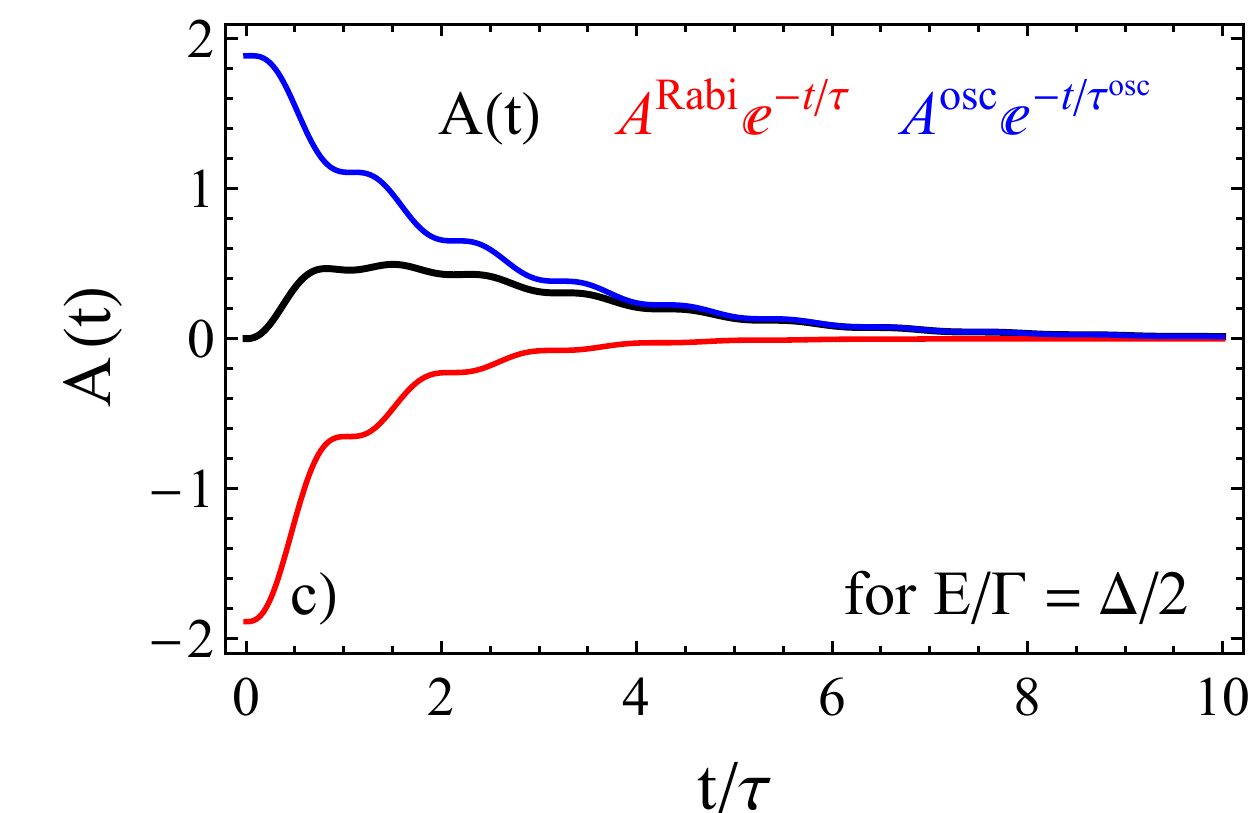}
\caption{a) $A(E,t)$ as a function of energy $E$ for selected times $t$.
b) and c) The time evolution of $A(E,t)$ and its components for $E = 0$ and $E = E_-$, respectively.
The other parameters are the same as in Fig.~3.} \label{fig4}
\end{figure}

For the strong dot-dot coupling, $t_{12} \gg \Gamma$, one can treat 2QD as an artificial molecule and observe the time evolution of the transmission from the single peak structure into the double peak structure, see Fig.~3. The transmission $T(E,t)$ is expressed as
\begin{align}\label{eq:TEtlarge}
T(E,t) &= \sum_{\nu = \pm} [ T_{1QD}^{osc,\nu}(E,t) + T_{2QD}^{osc,\nu}(E,t) ] e^{- t / \tau^{osc}} \nonumber \\
&+ T_{2QD}^s(E) \; ,
\end{align}
where its components are
\begin{align}
&T_{1QD}^{osc,\nu}(E,t) = - \nu \frac{\Gamma}{2 \Delta} \left\{ \frac{\Gamma ( \mathcal{E} - \nu \Delta ) \cos [( \mathcal{E} + \nu \Delta / 2 ) t ]}{\mathcal{E}^2 + \Gamma^2} \right. \nonumber \\
&+ \left. \frac{( \Gamma^2 + \nu \mathcal{E} \Delta ) \sin [( \mathcal{E} + \nu \Delta / 2 ) t ]}{\mathcal{E}^2 + \Gamma^2} \right\} \; , \\
&T_{2QD}^{osc,\nu}(E,t) = \nu \frac{\Gamma}{4 \Delta} \left\{ \frac{ 2 \Gamma \mathcal{E} \cos [( \mathcal{E} + \nu \Delta / 2 ) t ]}{( \mathcal{E} + \nu \Delta / 2 )^2 + \Gamma^2 / 4} \right. \nonumber \\
&+ \left. \frac{( \Gamma^2 + 2 \nu \Delta \mathcal{E} + \Delta^2 ) \sin [( \mathcal{E} + \nu \Delta / 2 ) t ]}{( \mathcal{E} + \nu \Delta / 2 )^2 + \Gamma^2 / 4} \right\} \; , \\
&T_{2QD}^s (E) = - \frac{1}{2 \Delta} \sum_{\nu = \pm} \frac{\nu \Gamma^2 \mathcal{E}}{( \mathcal{E} + \nu \Delta / 2 )^2 + \Gamma^2 / 4} \; . \label{eq:Tcomlarge}
\end{align}
Here, $\Delta = ( 4 t_{12}^2 - \Gamma^2 )^{1/2}$ denotes a separation between the bonding and anti-bonding energy levels, $E_{\pm} = \epsilon \mp \Delta / 2$.
Now, in the stationary limit $T_{2DQ}^s(E)$ has two asymmetric resonance peaks at $E_{\pm}$ with the broadening $\Gamma / 2$.
It is in contrast to the weak coupling case, Eq.~(\ref{eq:t2qds}), where two conducting channels have different broadenings $\Gamma_\pm$ but the resonance is at $E = \epsilon$.
Relaxation processes are different for the strong dot-dot coupling. The transmission evolves with a single relaxation time $\tau^{osc} \equiv 2 / \Gamma$ which describes coherent electron transfers to the external electrodes.
Now, the dot-electrode charge oscillations are more pronounced, which lead to a change in the sign of the transmission and consequently, in a change in the direction of the current flow, see Fig.~3.
The strongest changes of $T(E,t)$ one can observe for $E = 0$, i.e. in the valley between bonding and anti-bonding states when destructive interference takes place (in contrast with the previous case for the weak dot-dot coupling).
It is clear that at $E = 0$ the major changes are seen in the 1QD components, $T_{1QD}^{osc,\pm}(t) = [ \cos ( t \Delta / 2 ) - ( \Gamma / \Delta ) \sin ( t \Delta / 2 )] / 2$, while bonding and anti-bonding states also contribute equally, $T_{2QD}^{osc,\pm}(t) = ( \Gamma / \Delta ) \sin ( t \Delta / 2 )$, but play a minor role, see Fig.~3b).
Formation of the anti-bonding state at $E = E_- = \epsilon + \Delta / 2$ is presented in Fig.~3c).
It is caused mainly by an exponential decay of the negative component $T_{2QD}^{osc,-}(t)$ while the other components give small oscillations.

In the spectral function $A(E,t)$ one finds two components:
\begin{equation}
A(E,t) = A^{osc}(E,t) e^{- t / \tau^{osc}} + A^{Rabi}(E,t) e^{- t / \tau} \; .
\end{equation}
Apart from the aforementioned charge oscillations there are also rapid Rabi oscillations caused by coherent electron transfer between the 2QD states, see Fig.~4.
These components are proportional to $\sin ( t \Delta / 2 )$ and $\cos ( t \Delta / 2 )$.
The frequency of the Rabi oscillations $\omega^{Rabi} = 2 \pi \Delta / 2$ can be tuned by changing dot-dot coupling $t_{12}$.
In general, the Rabi component $A^{Rabi}(E,t)$ gives a negative contribution to the spectral function $A(E,t)$ and vanish with a relaxation time $\tau \equiv 1 / \Gamma$, i.e. twice as quick as charge oscillation component $A^{osc}(E,t)$, see Fig.~4 and for details \textit{Appendix}.

\section{Conclusions and outlook}

In the paper, we studied a role of quantum interference in time formation of the Fano resonance in the 2QD-T system.
We assumed that at the initial time $t = 0$ the 1QD system, with a single QD coupled with the electrodes, was in a stationary state, and next, the second QD was suddenly side attached to the first QD.
Transient tunneling processes were analyzed, in particular the time evolution of transmission, $T(E,t)$ and local charge relaxation (described by the local density $A(E,t)$ which is related with entanglement).
We found two regimes with different dynamics depending on the inter-dot coupling with respect to the dot-electrode coupling.
For the weak dot-dot coupling formation  of the Fano anti-resonance in $T(E,t)$ is governed by two channel tunnelling processes with two characteristic relaxation times.
In contrast, for the strong dot-dot coupling $T(E,t)$ evolves with a single relaxation time to two peak structure which corresponds to resonance transport through bonding and anti-bonding states.
One founds large charge oscillations between the bonding/anti-bonding states and the electrodes, which can even change direction of the transient current.
Additionally, in the time evolution of $A(E,t)$ one observes interplay of rapid Rabi oscillations (between the bonding and the anti-bonding state) and charge oscillations to the electrodes.

It will be interesting to extend our time dependent studies of charging effects and entanglement, analyzing entropy flow and showing various thermalization processes caused by quantum coherence which one expects will be different in the weak and strong coupling regime.
Another fascinating perspective will be the analysis of the current-current correlations in the transient regime, e.g.: to determine the dominant microscopic relaxation processes at different time scales; to show how quantum interference affect short-time fluctuations.

\section*{Acknowledgements}

The research was financed by the National Science Centre, Poland - project number 2016/21/B/ST3/02160.

\appendix

\section{Appendix}
\setcounter{equation}{0}
\renewcommand\theequation{A.\arabic{equation}}

The Appendix gives the explicit form of the components of the density function $A(E,t)$.

\subsection{Weak dot-dot coupling}

\noindent The components are separated $A^g(E) = A_{1QD}^g(E) + A_{2QD}^g(E)$, $A^\nu(E) = A_{1QD}^\nu(E) + A_{2QD}^\nu(E)$ and $A^{osc,\nu}(E,t) = A_{1QD}^{osc,\nu}(E,t) + A_{2QD}^{osc,\nu}(E,t)$ to show their time evolution from the 1QD to the 2QD system.
These components are expressed as:
\begin{align}
&A^g_{1QD}(E) = - \frac{\Gamma_+ \Gamma_-}{2 \delta^2} \frac{4 \Gamma^2 + \Gamma_+^2 + \Gamma_-^2}{9 \Gamma^2 - \delta^2} \frac{\Gamma^2}{\mathcal{E}^2 + \Gamma^2} \; , \\
&A^g_{2QD}(E) = \frac{\Gamma}{2 \delta^2} \sum_{\nu = \pm} \frac{\Gamma_- \Gamma_+}{2 \Gamma + \Gamma_\nu} \frac{\Gamma_\nu^2}{\mathcal{E}^2 + \Gamma_\nu^2 / 4} \; , \\
&A_{1QD}^\nu(E) = \frac{\Gamma^2}{2 \delta^2} \frac{\Gamma_- \Gamma_+}{2 \Gamma + \Gamma_\nu} \frac{\Gamma_\nu}{\mathcal{E}^2 + \Gamma^2} \; , \\
&A_{2QD}^\nu(E) = - \frac{\Gamma^2}{2 \delta^2} \frac{\Gamma_- \Gamma_+}{2 \Gamma + \Gamma_\nu} \frac{\Gamma_\nu}{\mathcal{E}^2 + \Gamma_\nu^2 / 4} \; , \\
&A_{1QD}^{osc,\nu}(E,t) = \nu \frac{\Gamma}{\delta} \frac{\Gamma_- \Gamma_+ \Gamma_\nu}{9 \Gamma^2 - \delta^2} \frac{\Gamma \cos ( \mathcal{E} t ) - \mathcal{E} \sin ( \mathcal{E} t )}{\mathcal{E}^2 + \Gamma^2} \; , \\
&A_{2QD}^{osc,\nu}(E,t) = \frac{\Gamma^2}{2 \delta^2} \frac{\Gamma_- \Gamma_+}{2 \Gamma + \Gamma_\nu} \frac{\Gamma_\nu \cos ( \mathcal{E} t ) + 2 \mathcal{E} \sin ( \mathcal{E} t )}{\mathcal{E}^2 + \Gamma_\nu^2 / 4} \nonumber \\
&- \frac{\Gamma}{2 \delta^2} \frac{\Gamma_\nu \Gamma_{\bar{\nu}}^2}{2 \Gamma + \Gamma_{\bar{\nu}}} \frac{\Gamma_{\bar{\nu}} \cos ( \mathcal{E} t ) + 2 \mathcal{E} \sin ( \mathcal{E} t )}{\mathcal{E}^2 + \Gamma_{\bar{\nu}}^2 / 4} \; ,
\end{align}
where $\bar{\nu} \equiv - \nu$.

\subsection{Strong dot-dot coupling}

\noindent In the strong coupling limit the components corresponding to the Rabi oscillations and related with the oscillations to the electrodes, are separated $A^{Rabi}(E,t) = A_{1QD}^{Rabi}(E,t) + A_{2QD}^{Rabi}(E,t)$ and $A^{osc}(E,t) = A_{1QD}^{osc}(E,t) + A_{2QD}^{osc}(E,t)$.
Their components are given by:
\begin{align}
&A^{Rabi}_{1QD}(E,t) = \frac{\Gamma^2}{\Delta^2} \frac{\Gamma^2 + \Delta^2}{9 \Gamma^2 + \Delta^2} \Big[ \frac{3 \Gamma^2 - \Delta^2}{\mathcal{E}^2 + \Gamma^2} \nonumber \\
&- \frac{( 3 \Gamma^2 + \Delta^2 ) \cos ( \Delta t ) + 2 \Gamma \Delta \sin ( \Delta t )}{\mathcal{E}^2 + \Gamma^2} \Big] \; , \\
&A^{Rabi}_{2QD}(E,t) = \frac{\Gamma^2}{2 \Delta^2} \frac{\Gamma^2 + \Delta^2}{9 \Gamma^2 + \Delta^2} \times \nonumber \\
&\sum_{\nu = \pm} \Big\{ \frac{- 3 \Gamma^2 - 4 \nu \Delta \mathcal{E} - 3 \Delta^2}{( \mathcal{E} + \nu \Delta / 2 )^2 + \Gamma^2 / 4} 
- \nu \frac{2 \Gamma ( 3 \mathcal{E} + 2 \nu \Delta ) \sin ( \Delta t )}{( \mathcal{E} + \nu \Delta / 2 )^2 + \Gamma^2 / 4} \nonumber \\
&+ \frac{( 3 \Gamma^2 - 2 \nu \Delta \mathcal{E} - \Delta^2 ) \cos ( \Delta t )}{( \mathcal{E} + \nu \Delta / 2 )^2 + \Gamma^2 / 4} \Big\} \; , \\
&A^{osc}_{1QD}(E,t) = - \frac{\Gamma}{\Delta} \frac{\Gamma^2 + \Delta^2}{9 \Gamma^2 + \Delta^2} \times \nonumber \\
&\sum_{\nu = \pm} \nu \Big\{ \frac{\Gamma ( \mathcal{E} - \nu \Delta ) \cos [( \mathcal{E} + \nu \Delta / 2 ) t]}{\mathcal{E}^2 + \Gamma^2} \nonumber \\
&+ \frac{( \Gamma^2 + \nu \Delta \mathcal{E} ) \sin [( \mathcal{E} + \nu \Delta / 2 ) t ]}{\mathcal{E}^2 + \Gamma^2} \Big\} \;, \\
&A^{osc}_{2QD}(E,t) = \frac{\Gamma}{2 \Delta^2} \frac{\Gamma^2 + \Delta^2}{9 \Gamma^2 + \Delta^2} \times \nonumber \\
&\sum_{\nu = \pm} \Big\{ \frac{\Gamma ( 3 \Gamma^2 + 4 \nu \Delta \mathcal{E} + 3 \Delta^2 ) \cos [( \mathcal{E} + \nu \Delta / 2 ) t ]}{( \mathcal{E} + \nu \Delta / 2 )^2 + \Gamma^2 / 4} \nonumber \\
&+ \nu \frac{[ \Gamma^2 \Delta + 2 \nu ( 3 \Gamma^2 + \Delta^2 ) \mathcal{E} + \Delta^3 ] \sin [( \mathcal{E} + \nu \Delta / 2 ) t ]}{( \mathcal{E} + \nu \Delta / 2 )^2 + \Gamma^2 / 4} \nonumber \\
&- \frac{\Gamma ( 3 \Gamma^2 - 2 \nu \Delta \mathcal{E} - \Delta^2 ) \cos [( \mathcal{E} - \nu \Delta / 2 ) t ]}{( \mathcal{E} + \nu \Delta / 2 )^2 + \Gamma^2 / 4} \nonumber \\
&- \frac{2 \Gamma^2 ( 3 \mathcal{E} + 2 \nu \Delta ) \sin [( \mathcal{E} - \nu \Delta / 2 ) t ]}{( \mathcal{E} + \nu \Delta / 2 )^2 + \Gamma^2 / 4} \Big\} \; .
\end{align}
Notice that $A_{1QD}^{osc}(E,t)$ and $A_{2QD}^{osc}(E,t)$ have two components related with charge oscillations to the bonding and anti-bonding states.


\begin{thebibliography}{99}

\bibitem{gruson2016}
\href{https://doi.org/10.1126/science.aah5188}
{V.~Gruson, L.~Barreau, \'{A}.~Jim\'{e}nez-Galan, F.~Risoud, J.~Caillat, A.~Maquet, B.~Carr\'{e}, F.~Lepetit, J.-F.~Hergott, T.~Ruchon, L.~Argenti, R.~Ta\"{\i}eb, F.~Mart\'{\i}n, P.~Sali\`{e}res, Science 354 (2016) 734}.

\bibitem{kaldun2016}
\href{https://doi.org/10.1126/science.aah6972}
{A.~Kaldun, A.~Bl\"{a}ttermann, V.~Stoo{\ss}, S.~Donsa, H.~Wei, R.~Pazourek, S.~Nagele, C.~Ott, C.D.~Lin, J.~Burgd\"{o}rfer, T.~Pfeifer, Science 354 (2016) 738}.

\bibitem{gores2000}
\href{https://doi.org/10.1103/PhysRevB.62.2188}
{J.~G\"{o}res, D.~Goldhaber-Gordon, S.~Heemeyer, M.A.~Kastner, H.~Shtrikman, D.~Mahalu, U.~Meirav, Phys. Rev. B 62 (2000) 2188}.

\bibitem{sasaki2009}
\href{https://doi.org/10.1103/PhysRevLett.103.266806}
{S.~Sasaki, H.~Tamura, T.~Akazaki, T.~Fujisawa, Phys. Rev. Lett. 103 (2009) 266806.}

\bibitem{bulka2005}
\href{https://doi.org/10.12693/APhysPolA.108.555}
{B.R.~Bu{\l}ka, P.~Stefa\'{n}ski and A.~Tagliacozzo, Acta Phys. Pol. A 108 (2005) 555.}

\bibitem{zitko2010}
\href{https://doi.org/10.1103/PhysRevB.81.115316}
{R.~\v{Z}itko, Phys. Rev. B 81 (2010) 115316.}

\bibitem{bulka2001}
\href{https://doi.org/10.1103/PhysRevLett.86.5128}
{B.R.~Bu{\l}ka, P.~Stefa{\'n}ski, Phys. Rev. Lett. 86 (2001) 5128.}

\bibitem{kang2001}
\href{https://doi.org/10.1103/PhysRevB.63.113304}
{K.~Kang, S.Y.~Cho, J.-J.~Kim, S.-C.~Shin, Phys. Rev. B 63 (2001) 113304.}

\bibitem{stefanski2004}
\href{https://doi.org/10.1103/PhysRevLett.93.186805}
{P.~Stefa{\'n}ski, A.~Tagliacozzo, B.R.~Bu{\l}ka, Phys. Rev. Lett. 93 (2004) 186805.}

\bibitem{pan2009}
\href{https://doi.org/10.1088/0953-8984/21/26/265501}
{H.~Pan and Y.~Zhao, J. Phys.: Condens. Matter 21 (2009) 265501}.

\bibitem{kwapinski2014}
\href{https://doi.org/10.1016/j.physe.2014.06.006}
{T.~Kwapi\'{n}ski, R.~Taranko, Physica E 63 (2014) 241}.

\bibitem{haug2008}
\href{https://doi.org/10.1007/978-3-540-73564-9}
{H.J.W.~Haug, A.-P.~Jauho, Quantum Kinetics in Transport and optics of Semiconductors (Springer, Berlin Heidelberg, 2008) p.~186}.

\bibitem{schaller2014}
\href{https://doi.org/10.1007/978-3-319-03877-3}
{G.~Schaller, Open Quantum Systems Far from Equilibrium, Lecture Notes in Physics 881 (Springer, Switzerland, 2014) p.~54}.

\bibitem{amico2008}
\href{https://doi.org/10.1103/RevModPhys.80.517}
{L.~Amico, R.~Fazio, A.~Osterloh, V.~Vedral, Rev. Mod. Phys. 80 (2008) 517}.

\bibitem{laflorencie2016}
\href{https://doi.org/10.1016/j.physrep.2016.06.008}
{N.~Laflorencie, Phys. Rep. 643 (2016) 1}.

\end{thebibliography}
\end{document}